# Surface and interface structure of quasi-free standing graphene on SiC

C Melios [1,2], S Spencer[1], A Shard[1], W Strupiński[3], S R P Silva[2] and O Kazakova[1]*

[1]National Physical Laboratory, Teddington, TW11 0LW, United Kingdom

[2]Advanced Technology Institute, University of Surrey, Guildford, Surrey, GU2 7XH, UK

[3]Institute of Electronic Materials Technology, Wólczyńska 133, 01-919 Warsaw, Poland

* E-mail: olga.kazakova@npl.co.uk

We perform local nanoscale studies of the surface and interface structure of hydrogen intercalated graphene on 4$H$-SiC(1000). In particular, we show that intercalation of the interfacial layer results in the formation of quasi-free standing one layer graphene (QFS 1LG) with change in the carrier type from n- to p-type, accompanied by a more than four times increase in carrier mobility. We demonstrate that surface enhanced Raman scattering (SERS) reveals the enhanced Raman signal of Si-H stretching mode, which is the direct proof of successful intercalation. Furthermore, the appearance of D, D +D' as well as C-H peaks for the quasi-free standing two layer graphene (QFS 2LG) suggests that hydrogen also penetrates in between the graphene layers to locally form C-H sp$^3$ defects that decrease the mobility. Thus, SERS provides a quick and reliable technique to investigate the interface structure of graphene which is in general not accessible by other conventional methods. Our findings are further confirmed by Kelvin probe force microscopy and X-ray photoelectron spectroscopy.

1. **Introduction**

The unique properties of graphene make it attractive for the use in nanoelectronics[1], among many other applications[2]. In particular, epitaxial graphene has shown great potential for integration in large-scale production due to its compatibility with CMOS fabrication process. Regardless of its exceptional electronic and structural properties, the vision of integrating graphene into nanoelectronic devices such as high-speed transistors[3], ring oscillators[4], integrated circuits[5] and many more depends on the capability to grow large-scale, uniform graphene of outstanding quality.

In this work, chemical vapor deposition (CVD) is used to grow graphene on the Si(0001) face of the 4*H*-SiC substrate by propane decomposition[6]. Despite the reproducible control of the growth process, the resulting graphene suffers from strong electron doping and limited mobility, due to the formation of the interfacial layer (IFL, also known as buffer layer), which is a carbon layer covalently bonded to the SiC substrate[7–9]. Several groups demonstrated that hydrogen intercalation of epitaxial graphene can decouple the IFL and convert it into a quasi-free standing graphene (QFSG), enhancing significantly the charge carrier mobility and even reversing the carrier type[7,8,10,11]. This intercalation process has resulted in high current saturation and improved intrinsic cut-off frequency graphene field-effect transistors[12], but important questions regarding the underlying physics of the intercalation process still remain unanswered.

Until now, X-ray photoelectron spectroscopy (XPS) has been used to characterize the surface composition and interface passivation of graphene samples globally[13,14], whereas Raman spectroscopy is typically used for characterization of its layer structure and doping on a local scale[8,15]. Although these techniques can be used in parallel to provide understanding of the complementary chemical and structural properties of graphene, they generally lack the spatial resolution (i.e. XPS) and chemical sensitivity to certain species, for example light atoms such as hydrogen (i.e. XPS and Raman spectroscopy in its standard implementation). In contrast to the standard confocal Raman technique[16], surface enhanced Raman scattering (SERS) can provide the spatial resolution of Raman spectroscopy, while accompanied by enhanced molecular sensitivity. SERS employs metal nanoparticles (MNPs) or fabricated patterned nanostructures on the surface of the sample to induce signal enhancement[17]. One of the main advantages of this method is its simplicity and straightforward operation, which is, however, somewhat lessened by lacks of the consistent control of the MNP size and, therefore the precise knowledge of scattering enhancement. In this work, we will explore the surface, interlayer (between graphene layers) and interface (between SiC and graphene) structure of different types of graphene samples using a combination of surface sensitive techniques such as: i) SERS to probe the underlying structure of the graphene layers and to reveal the presence of Si-H and C-H bonds on a local scale, ii) XPS to provide structural and chemical information of the as-grown and hydrogen intercalated graphene samples on a global scale, iii) Kelvin probe force microscopy (KPFM) to construct a detailed image of the surface potential of the layer structure.

Here, we consider three types of samples, namely: i) as-grown one layer (1LG) used as a reference sample; ii) quasi-free standing one layer graphene (QFS 1LG) obtained as a result of hydrogen intercalation of IFL; iii) quasi-free standing two layer graphene (QFS 2LG) obtained as a result of hydrogen intercalation of 1LG. By demonstrating the enhanced Raman signal of the Si-H bonds in both intercalated samples (QFS 1 and 2LG) we prove that hydrogen passivates the Si atoms of the SiC substrate, effectively decoupling the IFL from the substrate. Furthermore, by establishing that the C-H peak is present only in the QFS 2LG, we show that in this sample a relatively small fraction of hydrogen atoms not only passivates the Si atoms of the SiC substrate, but also penetrates between the graphene layers to locally form $sp^3$ C-H bond defects as proved by the strong D-peak in Raman.

We demonstrate that SERS is a quick and reliable method to establish the successful intercalation of the graphene and to provide complete electronic and structural information on both the global and local scales, which is not easily obtainable using conventional Raman spectroscopy.

## 2. Experimental section

*2.1. Sample Growth:*

The graphene samples were grown by CVD method at 1600 °C on semi-insulating on-axis oriented 4*H*-SiC (0001) substrates, under an argon laminar flow in an Aixtron VP508 hot-wall reactor. The 10×10 mm$^2$ size substrates (Cree) were cut out from 4" wafer and etched in hydrogen at 1600 °C prior to the growth process. The graphene growth was controlled by the Ar pressure, linear flow velocity and reactor temperature. This process relies critically on the generation of dynamic flow conditions in the reactor, which controls the Si sublimation rate and allows the mass transport of hydrocarbons to the SiC substrate. By tuning the value of the Reynolds number, a thick enough Ar boundary layer is formed which prevents Si sublimation and allows the diffusion of hydrocarbons to the SiC surface. This results in epitaxial growth of graphene on the SiC surface[6]. The intercalation of hydrogen was achieved by annealing the sample in hydrogen at temperature of 700 -1100 °C (depending on the sample) and reactor pressure of 900 mbar. Cooling down in H$_2$ atmosphere prevents hydrogen atoms trapped between graphene and substrate to escape. Prior to unloading the sample, the process gas was changed back to argon[8].

*2.2. Raman spectroscopy:*

Au particles in powder form of diameter ~0.8-1.5 μm (Alfa Aesar, 99.96%, CAS: 7440-57-5, catalog number: 39817) were randomly deposited on the sample surface using dry N$_2$. The dry deposition of the Au particles allows to re-use the sample. Raman maps of 30×30 μm$^2$ were obtained using a Horiba Jobin-Yvon HR800 system in order to investigate the structure of graphene samples. The 632.8 nm wavelength laser was focused through a 50× objective lens. The spectral resolution was ~1 cm$^{-1}$. The Raman spectra (without any MNP) were initially obtained for a reference SiC substrate, which was used as a control spectrum. The Raman maps were constructed by mapping the Si-H and 2D peak intensity and shift of 1936 individual spectra.

The 2D-peak is a dispersive, second order, inter-valley Raman scattering process, which originates from the in-plane (breathing-like) vibrations of carbon atoms. This process takes place at the K point of the first Brillouin zone, where the incident photon generates an electron-hole pair, which is inelasticity scattered by a phonon to the K' point. Because of energy and momentum conservation, the electron is being scattered back to the K point which then recombines to a hole[18,19]. This peak is of important significance, as it allows for accurately determination of graphene layers by measuring its position, FWHM and line-shape. In addition to the layer dependence of 2D-peak, 2D can be used as a measure of doping in graphene[20]. The position of 2D-peak highly depends on the doping and mechanical strain in graphene, where both mechanisms result in changes of the lattice constant[18,20,21].

The D-peak, often referred as a disorder peak, is a second order scattering process, where a defect is required for its activation. A photogenerated electron-hole pair scatters in-elastically by a phonon from K to K' point, which then scatters back to K elastically. Since a defect (such as $sp^3$ bonds or discontinuity) in the graphene lattice is required for this back-scattering process to occurs, the ratio of D/G intensity is a measure of defects in the lattice[19].

*2.3. Surface potential measurements:*

NT-MDT Aura scanning probe microscope was used for the surface potential measurements of the graphene samples in vacuum ($1 \times 10^{-6}$ mbar). Doped Si PFQNE-AL tips (provided by Bruker) with spring constant of k≈0.4–1.2 N m$^{-1}$ were used. In frequency modulated-Kelvin probe force microscopy (FM-KPFM), the cantilever oscillates at its mechanical frequency $f_0$≈300 kHz, while an AC voltage of a significantly lower frequency $f_{mod}$≈3 kHz is also applied, resulting in a frequency shift.

The $f_0 \pm f_{mod}$ side lobes (monitored by a PID feedback loop) generated by the frequency shift are minimized by applying a DC compensation voltage. By measuring this DC voltage at each pixel, a surface potential map [i.e. contact potential difference ($U_{CPD}$)] is constructed[22,23]. Because in FM-KPFM the force gradient is being detected, a spatial resolution of <20 nm can be achieved, which is limited only by the tip apex diameter, allowing determination of the number of graphene layers with great accuracy[24–27].

*2.4. X-ray photoelectron spectroscopy:*

X-ray photoelectron spectroscopy (XPS) is a surface sensitive technique carried out in ultra-high vacuum where incident x-ray photons cause emission of core-level electrons from a sample. Spectra are obtained which can be used to provide quantitative chemical composition and chemical state information. A Kratos Axis Ultra DLD x-ray photoelectron spectrometer was used. Samples were held in place on the sample stage by small copper clamps providing an electrical path between the top surface and earth. Carbon 1s spectra were acquired at an emission angle of 0° to the surface normal using a monochromated aluminium x-ray source operated at 15 kV, 5mA emission. Each analysis area was approximately 700 x 300 μm, the information depth being less than 10 nm. Analysis conditions were 20 eV pass energy, 0.1 eV steps, 0.5 sec dwell per step and 6 sweeps. The charge neutraliser was switched on for the spectra, however repeat spectra were also acquired with it switched off. Both sets of spectra appeared identical, indicating that there were no electrical charging effects. These spectra were then processed in CasaXPS software to obtain chemical state information for carbon. Seven components were fitted to the spectra representing carbon in the following states, SiC, $sp^2$, S1, S2, C-O, C=O and O=C-O. The $sp^2$ peaks were fitted with a Doniach-Sunjic asymmetric line-shape (α=0,01) due to the metallic nature of graphene[9], with 40% Gaussian contribution. The remaining peaks were fitted using Gaussian (25%)-Lorentzian fits.

### 3. Results and discussion

We first employ KPFM in order to construct a layer-resolved image of the graphene structure. It has been shown, that KPFM is a powerful technique that enables mapping of the surface potential (SP) variations between different graphene layers, thus constructing a nanometer resolution map of the SP[28]. Here, FM-KPFM is used in a single pass mode, where the topography and the SP of the sample are measured simultaneously at different frequencies. Since KPFM measures the difference between the tip ($\Phi_{Tip}$) and the sample ($\Phi_{sample}$) work functions, using $\Phi_{sample} = \Phi_{Tip} - eU_{CPD}$, the resulted map displays the variations in the local work function and indirectly the carrier concentration of the sample. In this experiment, the samples were annealed and measured in vacuum ($\sim 1 \times 10^{-6}$ mbar) to reveal their pristine work function and avoid any undesirable environmental doping.

The SP map of the as-grown sample is displayed in Figure 1a. The SP map reveals SiC terraces covered by a continuous layer of 1LG (~80%). Furthermore, the edges are covered with 2 (~12%) and 3LG (~8%), exhibiting lower SP). Although Raman spectroscopy has previously been used for identification of number of graphene layers[16,18], the technique suffers from its limited spatial resolution. In this work, we correlate the Raman measurements with the information obtained by FM-KPFM to confidently assign the different contrasts to specific number of layers[28]. This is particularly useful in the case of intercalated graphene, where the graphene layers are rearranged. To quantify the SP measurements, the work function of the scanning probe was calibrated, and the work functions of 1, 2 and 3LG were measured to be $\Phi_{1LG} = 4.33\ eV$, $\Phi_{2LG} = 4.36\ eV$ and $\Phi_{3LG} = 4.40\ eV$, respectively.

For the QFS 1LG, first a complete IFL was grown on the SiC substrate. As it is not possible to entirely avoid the nucleation growth, certain areas promoting graphene growth, such as terrace edges, might be already covered by 1LG. Upon annealing of the IFL in hydrogen environment, the hydrogen is forced between the IFL and substrate. This results in the passivation of Si dangling bonds and creates Si-H bonds, transforming the IFL to QFS 1LG. Upon this, the conduction switched to p-type and the mobility was enhanced to $\mu_h \approx 3900$ cm$^2$ (V·s)$^{-1}$, compared to $\mu_e \approx 860$ cm$^2$ (V·s)$^{-1}$ for the as-grown sample (the carrier concentration and mobility of the samples obtained by transport measurements in the van der Pauw geometry are shown in Table 1). In this case, the SP map of QFS 1LG (Figure 1b) shows SiC terraces covered with continuous QFS 1LG (~88%), and the terrace edges decorated with 2 (~3%) and 3LG (~9% as indicated by the darkest contrast). Due to the nanometer resolution of KPFM, it is possible to use it as a tool to assess the intercalation uniformity. The surface potential in figure 1b shows excellent uniformity, with no spatial variations, indicating uniform intercalation of the QFS 1LG. In this case the work function values are $\Phi_{1LG} = 4.66\ eV$, $\Phi_{2LG} = 4.75\ eV$ and $\Phi_{3LG} = 4.88\ eV$ for 1, 2 and 3LG, respectively. The significant increase in work function is another indication that the carriers switch from electrons to holes following intercalation.

For the growth of the QFS 2LG sample, initially the SiC substrate was graphitized until 1LG was formed, which was then intercalated to form QFS 2LG, i.e. in a similar way to the one described above. In this case, the 1LG has been transformed into 2LG and, in general, the as-grown layers (*n*) have been transformed into (*n* + 1)LG, as followed by conversion of the IFL into 1LG. The SP map of QFS 2LG in Figure 1c, shows terraces covered with free-standing 2LG (~86%), while 3 (~9%) and 4LG (~5%) appear at the terrace edges. Despite the successful intercalation of the graphene, some bubble-like structures appear in the topography (Supplementary information S1). These structures contain trapped excess hydrogen, which lifts up the graphene layers. It is however not possible to distinguish whether the hydrogen is located between the SiC substrate and the first graphene layer or in between the graphene layers. Work function values were calibrated to be $\Phi_{2LG} = 5.09\ eV$, $\Phi_{3LG} = 5.21\ eV$ and $\Phi_{4LG} = 5.24\ eV$ for 1, 2 and 3LG, respectively, which again shows that the Fermi level has crossed the Dirac point to hole conduction (p≈1.1×10$^{13}$ cm$^{-2}$ and μ$_h$≈2600 cm$^2$ (V·s)$^{-1}$).

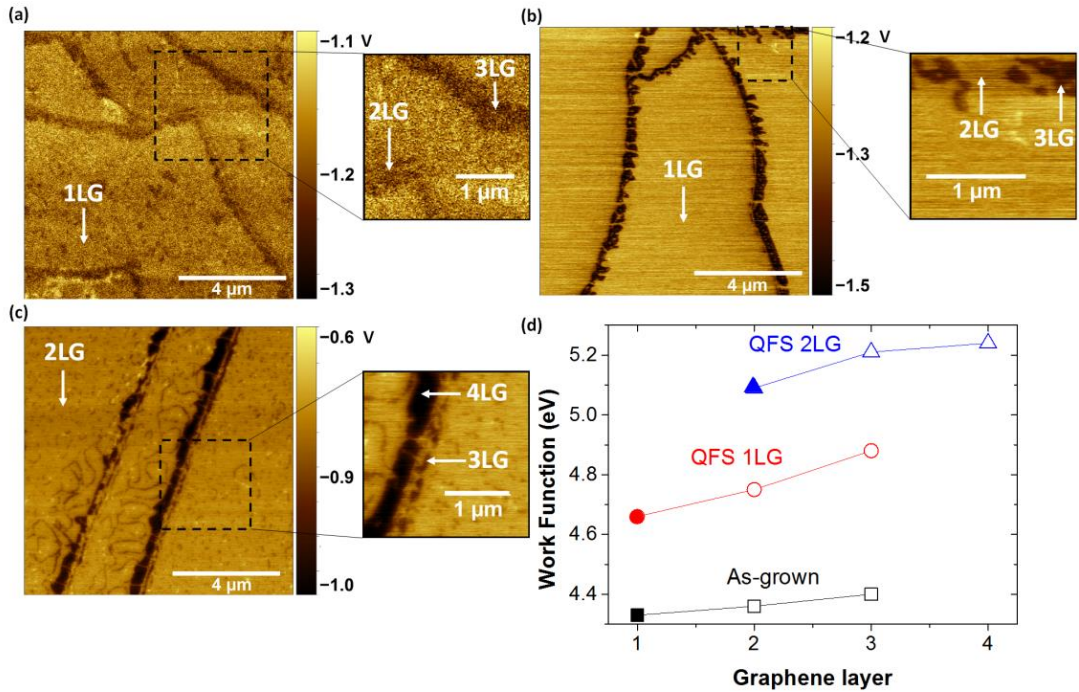

**Figure 1:** Surface potential maps in vacuum for the (a) as-grown sample, showing terraces covered by continuous 1LG and individual 2LG islands as well as elongated 2-3LG domains at the edges, (b) QFS 1LG, showing terraces covered by quasi-free standing 1LG as well as 2-3LG edges and (c) QFS 2LG, showing terraces covered by free standing 2LG and 3-4LG edges. The inset images display a zoom-in region of the enclosed areas. (d) Calibrated work function for the different graphene layers measured in vacuum for as-grown (black), QFS 1LG (red) and QFS 2LG (blue). Solid and open symbols correspond to measurements on terraces and edges, respectively.

**Table 1:** Carrier concentration and mobility of the samples used in the experiment.

| Sample | Carrier type | Carrier concentration ($cm^{-2}$) | Carrier mobility ($cm^2$ $(V·s)^{-1}$) |
|---|---|---|---|
| **As-grown 1LG** | Electrons | $-5.6 \times 10^{12}$ | 860 |
| **Quasi-free standing 1LG** | Holes | $+1.3 \times 10^{13}$ | 3900 |
| **Quasi-free standing 2LG** | Holes | $+1.1 \times 10^{13}$ | 2260 |

For the investigation of graphene chemical composition, X-ray photoelectron spectroscopy was employed in order to monitor the evolution of the C1s spectrum of the as-grown and intercalated samples (Figure 2). Seven fitted components can resolve the C1s core-level spectrum of the as-grown sample in Figure 2a, four of them (SiC, $sp^2$, S1 and S2 peaks) are of particular interest. Additionally, in all the graphene samples the carbon-oxygen peaks (286.64-289 eV), ~1.5% area, are present due to contamination of the graphene surface. The SiC peak at 283.76 eV is attributed to the carbon atoms of the SiC substrate, while the $sp^2$ peak at 284.54 eV is attributed to the graphene layer ($sp^2$ hybridized carbon). The remaining two main components S1 (284.85 eV) and S2 (285.1 eV) are attributed to the IFL[9,29,30]. Several photoelectron spectroscopy studies have demonstrated that the S1 component is associated with atoms in the IFL, which are strongly interacting with the underlying substrate through dangling bonds, whereas the S2 component is associated with the remaining $sp^2$ C atoms in the IFL[9,29,30]. The ration of S2:S1≈1.5. Following intercalation of the IFL, the C1s spectrum of the QFS 1LG in Figure 2b shows significant separation of the sp$^2$ and SiC peaks due to the shift of the latter to lower binding energies (282.27 eV). This shift is attributed to the band bending at the interface between graphene and SiC and is related to the SiC polytype polarization of the substrate[31,32]. Furthermore, the S1 and S2 components of the IFL are not present in the C1s spectrum of the QFS 1LG, proving the conversion of the IFL to quasi-freestanding graphene. The C1s spectrum of the QFS 2LG in Figure 2c shows again well-defined separation of the sp$^2$ and SiC peaks. In this case, the energy difference between the sp$^2$ and SiC peaks remains the same as in the QFS 1LG (~1.92 eV), proving that the band bending is related to the SiC substrate properties (polarization). Additionally, the S1 and S2 components are not present in this spectrum, signifying the quasi-free standing nature of the sample. A significant observation in this case is the increase of the sp$^2$:SiC area ratio of the QFS 2LG compared to the QFS 1LG (1.43% compared to 1.05%, respectively), which is the result of thicker graphene layers.

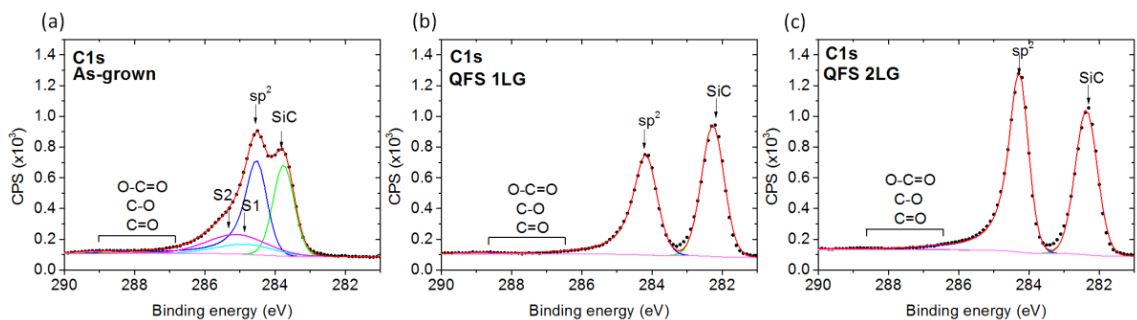

**Figure 2:** XPS measurements of (a) as-grown, (b) QFS 1LG and (c) QFS 2LG samples. The data are shown as points and fitted components as solid colored lines (red line is the overall envelope).

Despite XPS being a powerful technique for studying the graphene structure, the lack of spatial sensitivity and the requirement of high vacuum make it a rather complicated tool for quick graphene characterization. To overcome this limitations, SERS can be used to investigate the structure of graphene. For the investigation of the SERS on the graphene samples, Au particles in powder form of diameter ~0.8-1.5 μm were randomly deposited on the sample surface using dry $N_2$. The distribution of the Au particles coverage was assessed using scanning electron microscopy and the average diameter of 1.34 μm was derived (see Figure 3 a and b). The laser was focused through a 50× objective lens directly on top of the Au particle, which results in maximum signal enhancement. The control experiment showed that change of the laser focus through the whole diameter of the Au particle only affected the intensity of corresponding peaks, with maximum intensity achieved when the laser was focused on top of the particle[33]. We first compare Raman spectra for all studied graphene samples. Figure 3c shows the Raman spectrum measured on the bare SiC substrate to be used as a reference (black) and the SERS spectra of as-grown 1LG (red), QFS 1LG (green) and QFS 2LG (blue) obtained using Au MNPs (see experimental section for details). In all cases presented in Figure 3c, Raman spectra were obtained on terraces of the representative samples. To investigate the effects of scattering enhancement, the spectra were normalized to the 4$H$-SiC peak at ~777 cm$^{-1}$. The black spectrum shows a characteristic profile of the SiC substrate (this spectrum was measured using standard confocal Raman technique, subtracted spectra are presented in supplementary information S2). The graphene samples exhibit the main characteristic peaks, such as the G-peak (~1560 cm$^{-1}$, overlaid by the SiC peaks), 2D-peak (~2650 cm$^{-1}$) and D-peak (~1325 cm$^{-1}$).

Focusing at the 2D-peak of the as-grown graphene sample (red), the single symmetrical Lorenztian fit of the 2D-peak in Figure 3d is an indication of 1LG coverage[15]. It is important to note that due to the 50× objective used in this experiment, the spatial resolution of the Raman measurements is decreased and the layer assignment using fitting, shift and full-width-at-half-maximum (FWHM) must not be taken as absolute criteria for small areas (<1 μm). In this case the correlation with KPFM is proven to be an advantageous tool. An important observation is the lack of Si-H peak (~2130 cm$^{-1}$)[33], even when a MNP is present, signifying the absence of hydrogen bonded to the substrate, and the pristine nature of the graphene. Furthermore, the relatively low intensity of D-peak compare to G-peak indicates excellent quality graphene with no significant amount of defects present.

The green spectrum displayed in Figure 3c corresponds to QFS 1LG. The striking feature here, which distinguishes it from the as-grown sample, is the appearance of the peak at ~2130 cm$^{-1}$ corresponding to formation of the Si-H bond[33], a signature of the successful intercalation. Here, the advantage of SERS which enables the identification of Si-H peak and the confirmation of successful intercalation compared to standard Raman scattering is apparent. A comparison of spectra without MNP (normal Raman spectroscopy) is displayed in Figure 4 d and e, which will be discussed later in this paper. A noteworthy observation is the significant enhancement of the carrier mobility after intercalation, which is the result of excellent quality graphene. This is also reflected to the low intensity of the D-peak.

The blue spectrum in Figure 3c was taken on QFS 2LG. The 2D-peak in Figure 3f shows the typical line shape of *AB* stacked 2LG, which can be fitted with four Lorentzians[18,21,34]. The blue spectrum of QFS 2LG also exhibits a Si-H peak, confirming the successful intercalation of the sample. In addition to the Si-H peak, a broader low-intensity peak appeared at ~2915 cm$^{-1}$. This peak (inset of Figure 3c) can be deconvoluted by three components with peaks at ~2857, 2915 and 2969 cm$^{-1}$, with the last one corresponding to the second order D+D' peak of graphene, which is activated by the formation of defects[35]. Despite the decoupling of the graphene from the substrate, which enhanced the mobility dramatically compared to the as-grown sample (860 cm$^2$ (V·s)$^{-1}$), the introduction of defects in QFS 2LG is reflected by the lower carrier mobility (2260 cm$^2$ (V·s)$^{-1}$) when compared with the QFS 1LG (3900 cm$^2$ (V·s)$^{-1}$) (Table 1). The significant increase of D-peak is a clear indication of the defect formation in the graphene structure. In contrast with the formation of long-range structural disorder, observed by Hong et. al. in the graphene Raman spectrum where the D-peak is much broader [36], the sharpness of the D-peak in QFS 2LG is an indication of point defects in the graphene structure (vacancies and sp$^3$ bonded carbon). Previous studies have also shown that C-H vibrational modes occur in this range, resulting to the symmetric and asymmetric stretching vibration modes around ~2857 and 2915 cm$^{-1}$ [37–39]. The adsorption of hydrogen by the graphene lattice leads to the formation of local sp$^3$ hybridization of the graphene π* orbital, and thus to the increase of D and D+D' peaks as well as the formation of C-H vibration modes. The establishment of C-H vibrational modes can explain the formation of local carbon-hydrogen bonds between the layers of QFS 2LG. The absence of the C-H peak in the QFS 1LG spectrum suggests that hydrogen only bonds to the Si atoms of the SiC substrate, while in QFS 2LG, hydrogen also penetrates between graphene layers to form C-H bonds. The formation of C-H bonds was also previously demonstrated in hydrogenated graphene [14,39–41]. Hydrogenation brakes the C-C sp$^2$ bonds and forms C-H sp$^3$ bonds, giving rise to sharp D and D+D' peaks in the Raman spectrum [41]. Furthermore, C-H bonds were demonstrated in hydrogenated amorphous and diamond-like carbon [38]. Although SERS has been proven to be a powerful technique to probe the Si-H and C-H bonds in the intercalated graphene, the question of the hydrogen position (on top or in between of graphene layers) still remains unanswered. Even though we do not have a direct proof that hydrogen is placed on top of the graphene or between the layers (in the case of QFS 2LG), several observations point to the last scenario. Firstly, the enhanced D and D+D' peaks of the QFS 2LG compared to QFS 1LG in Figure 3c suggest that C-H bonds ($sp^3$ defects) only occur in the first case, where hydrogen penetrates between the layers, which is the result of the higher annealing temperature of QFS 2LG, compared to QFS 1LG. Furthermore, using Transmission Electron Microscopy and X-ray diffraction, Tokarczyk et. al.[8] demonstrated that the interlayer spacing between the graphene layers was increased following intercalation.

Considering the potential of integrating intercalated graphene in high-speed analogue electronics, it is important to comment on these findings and demonstrate that QFS 1LG shows superior quality and higher carrier mobility compared to QFS 2LG and therefore it should be the preferred material for such applications.

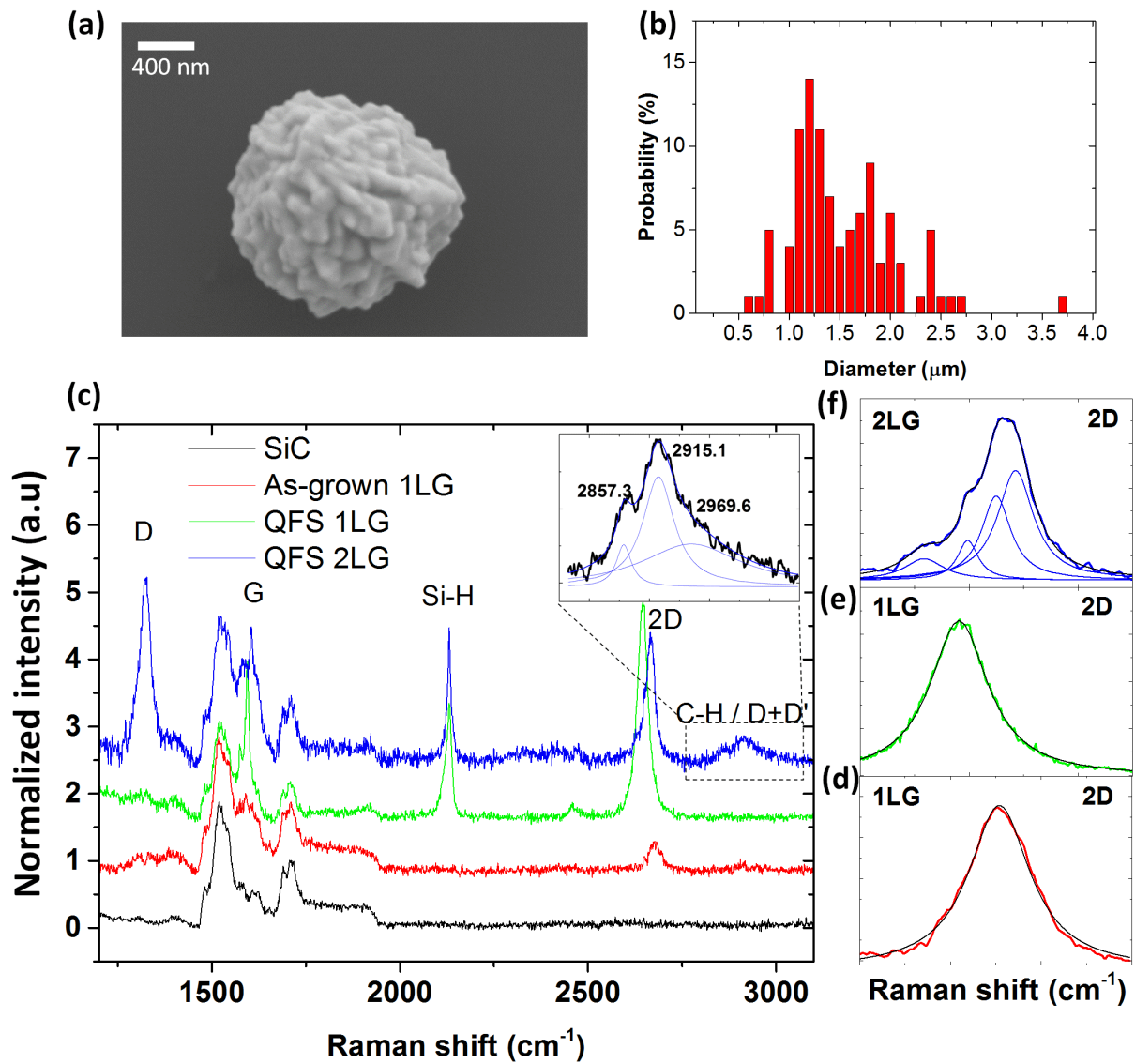

**Figure 3:** (a) SEM micrograph of a gold MNP deposited on graphene sample, (b) Statistical analysis of gold MNPs, showing median diameter of 1.34 μm, (c) SERS spectra measured on the bare SiC substrate as a reference (black), as-grown 1LG (red), QFS 1LG (green) and QFS 2LG (blue). The enhanced Si-H peak is an indication of successful passivation of the substrate Si atoms with hydrogen and the consequent intercalation of the graphene. The inset in top right shows the fitting of the peak at ~2916 cm$^{-1}$, which corresponds to the C-H bonds and D-D' graphene mode. (d-f) 2D fitted peaks of as-grown 1LG, QFS 1LG and QFS 2LG, respectively.

Next, we focus on detailed studies of QFS 1LG sample. For further investigation of the scattering enhancement of different size MNPs on different locations of the sample (terraces and edges), the Raman maps of QFS 1LG sample (Figure 4) were assembled by mapping the Si-H and 2D peak intensity and shift of 1936 individual spectra. Representative spectra were collected and displayed in Figure 4d,e. By analyzing the black 2D-peak, extracted from the terrace, the number of graphene layers in this area can be estimated. Based on the narrow and symmetrical 2D-peak of the black spectrum in Figure 4d (fitted by a single Lorentzian) it can be deduced that the SiC terraces are covered by 1LG. Additionally, variations of the 2D-peak shift are about 8 cm$^{-1}$ on the terrace (Figure 4c, enclosed dashed area and histogram inset), suggesting that the QFS 1LG experiences some local charge inhomogeneities and strain[19,42]. Furthermore, the individual spectra obtained in vicinity of MNPs on terraces are presented in Figure 4d. The symmetrical line-shape of the spectra indicates that the MNPs on terraces are positioned on 1LG. Despite that, it can be seen that the positions of some of the 2D-peaks with MNP are red-shifted (~6 cm$^{-1}$), i.e., for particles #2 and #7, compared to the peak without MNP. This can either suggest strained graphene induced by the Au MNPs[43] or more likely is due to the convolution effects of 1 and 2LG at the vicinity of terrace edges (MNPs #2 and #7 are positioned very close to edges). Due to the 50× objective used in this experiment, the laser spot size is ~1.02 μm. This decreases the spatial resolution of the mapping, which introduces peak convolution effects in areas near the terrace edges.

As discussed previously, terrace edges are typically covered with 2LG following intercalation. These areas can be seen in Figure 4c as blue elongated stripes, where the 2D-peak is blue-shifted. Individual spectra of 2LG at the terrace edges have been collected and shown in Figure 4e. The 2D-peak (olive) collected form the location far away from a MNP and marked with Edge 1 is blue-shifted compared to 1LG. Additionally, its broader and asymmetric line-shape signifies that the blue areas in Figure 4c are covered by 2LG (and possibly thicker graphene). Lastly, no graphene is observed in the area marked as SiC, which results in the absence of 2D-peak in Figure 4e.

Scattering enhancement depends highly on the MNP size[43]. To understand this dependence, one has to consider the classical Mie theory[44], i.e. an analytical solution of Maxwell's equations for spherical nanoparticles (which does not take into account any quantum or chemical information). In the case of a relative large particle but smaller than the wavelength of the incident wave, the sphere can be approximated as a dipole with size of the order of the incident wave. Here, the polarizability ($\alpha$) can be expressed in terms of the radius ($r$) and dielectric permittivity ($\varepsilon_{m(\omega)}$) of the MNP and surrounding medium ($\varepsilon_d$), respectively[45]:

$$\alpha = 4\pi r^3 \frac{\varepsilon_{m(\omega)} - \varepsilon_d}{\varepsilon_{m(\omega)} + 2\varepsilon_d}$$

When $\varepsilon_{m(\omega)} = -2\varepsilon_d$, the condition for localized plasmon resonant (LPR) is satisfied and the scattering cross-section ($k_{ext} = k_{sca} + k_{abs}$) exceeds the geometrical cross-section of the MNP. For maximum enhancement the absorption coefficient $k_{abs} = \frac{2\pi}{\lambda} Im(\alpha)$, needs to be minimum, while the scattering coefficient, $k_{sca} = \frac{1}{6\pi} \left(\frac{2\pi}{\lambda}\right)^4 |\alpha|^2$ needs to be maximum[45]. This means that $k_{sca}$ is highly dependent on the size, surrounding medium and incident wavelength. Despite the size of the particle is larger than the incident wavelength in our experiment (the large particles were chosen, so they will be optically visible), we are still able to achieve significant enhancement. As the theory predicts, larger particles result in higher scattering enhancement. This is also true in our experiments, where both Si-H and 2D peak exhibit different intensity enhancement for particles with different sizes (figure 4 a and b). Furthermore, the size dependence of the scattering enhancement in this experiment is reflected in Figure 4 d and e, where Si-H and 2D peaks are plotted for different size nanoparticles. The maximum enhancement of the Si-H Raman signal is found in the orange spectrum (MNP #5) of Figure 4, where the Si-H peak is enhanced by ~7 times compared to the Si-H peak of MNP #4 (MNP #5 is ~1.3 times larger than MNP #4, as measured optically). A statistical analysis of the MNP sizes was done using scanning electron microscopy (SEM), displayed in Figure 3 a and b.

We further focus at the Si-H peaks of the spectra collected near MNP #2, 7 (both on 1LG) and 5 (2LG), with a characteristic broad line shape and a shoulder appearing at ~2117 and 2128 cm$^{-1}$ for MNP #7 and 5, respectively, see insets in Figures 4 d, e. The reason for this asymmetric peak might be the formation of Si-H and Si-H$_2$ bonds, that vibrate in symmetric and asymmetric modes, generating a peak that can be deconvoluted into two components[33,46–49]. An alternative explanation is related to the different third-neighboring atoms of the SiC top surface terminations, where one of the Si-H peak components is associated with the cubic terminations of the Si lattice, while the other with its hexagonal terminations[50,51]. Furthermore, the terrace edges convolution might also be involved in the splitting of the Si-H peak. It is worth noting that the position of the Si-H peak of the 1LG at the terraces and the 2LG at the edges varies between 2117 cm$^{-1}$ and 2132 cm$^{-1}$. In this case, the Si-H bond is confined under thicker graphene, thus stiffening the vibrational frequency[33].

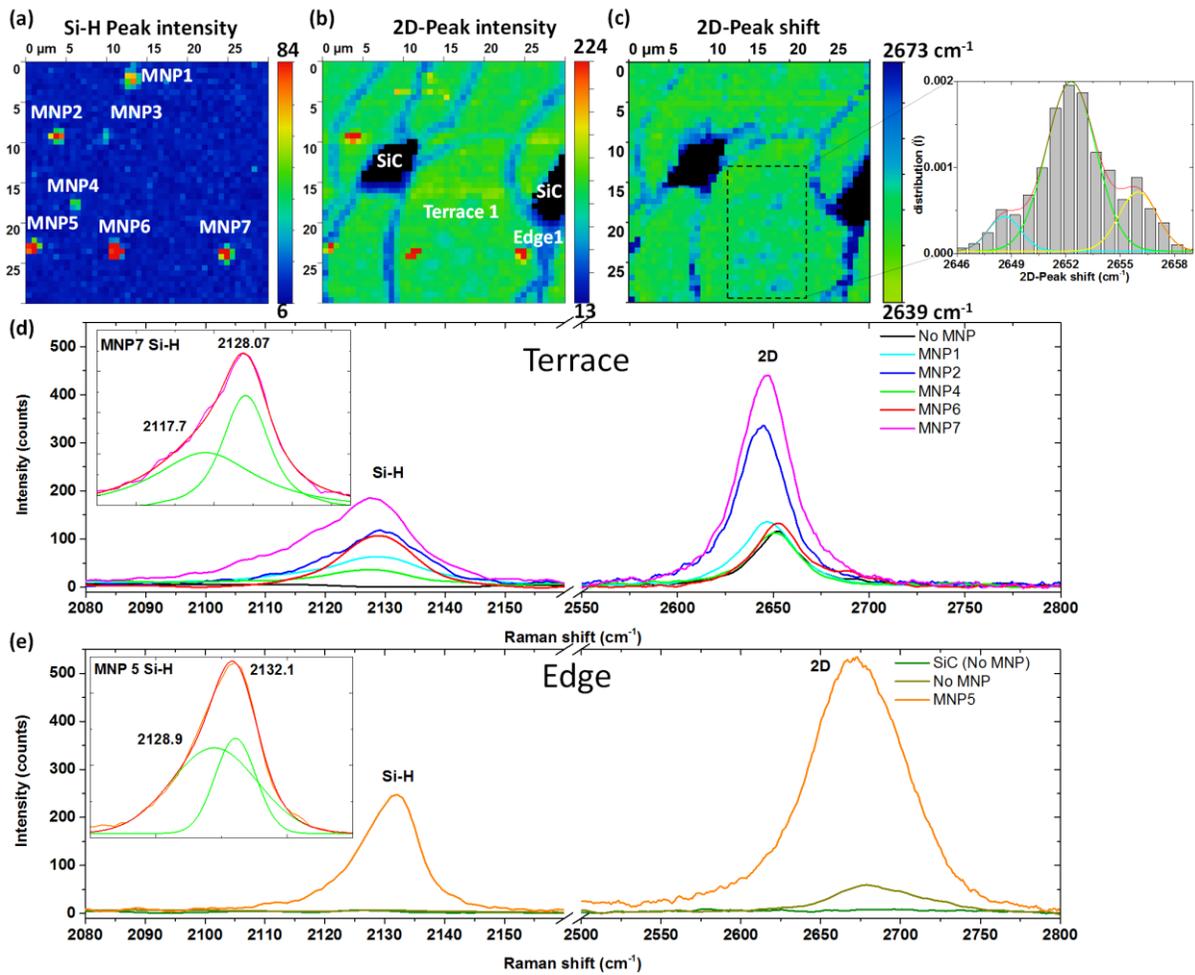

**Figure 4:** SERS maps of (a) Si-H peak intensity, (b) 2D-peak intensity and (c) 2D-peak shift of the QFS 1LG sample. The green areas of 2D-peak intensity and shift corresponds to 1LG, while the blue to 2LG.

The histogram inset in (c) shows the variations in the 2D-peak shift on the terraces. (d, e) Individual Raman spectra collected from the marked areas of the maps near specific MNPs on (d) terraces and (e) edges, showing the enhanced Si-H and 2D-peaks. The insets in (d, e) show the Si-H peak fitting for MNP #7 and MNP #5.

## 4. Conclusion

We investigated the interface structure of hydrogen intercalated CVD graphene on 4*H*-SiC(1000) using a combination of FM-KPFM, SERS and XPS techniques on local and global scales. Using XPS, we monitored the transformation of the C1s peak, constructing a global image of the evolution of the IFL to QFS 1LG and 2LG. Upon intercalation of the IFL, the S1 and S2 components were absent, signifying the transformation of the IFL to QFS 1LG. The intercalated 1LG, resulted in QFS 2LG, exhibiting increase of the total number $sp^2$ bonds. We also studied local structural and chemical properties of pristine and hydrogen intercalated graphene by means of SERS technique. Gold nanoparticles were deposited on the samples, in order to enhance the Raman signal of Si-H and C-H vibrational modes and study the effects of intercalation of graphene of different thicknesses. The as-grown sample demonstrated excellent quality of graphene structure, with no Si-H peak present, signifying that the IFL is covalently bonded to the substrate. Following hydrogen intercalation, the Si-H peak has appeared, proving that hydrogen passivated the Si atoms of the SiC substrate and converted the IFL to QFS 1LG. The intercalation process increases significantly the electronic properties and particularly the hole mobility, which is increased by more than 4 times compared to the as-grown sample. The use of SERS on the QFS 2LG revealed both the Si-H and the C-H peaks, suggesting that hydrogen not only passivates the Si atoms of the substrate, but also bonds to the graphene to locally form $sp^3$ type defects, resulting in significant increase of the D-peak. Despite the successful intercalation of QFS 2LG, which increased the hole mobility compared to the as-grown sample, the induced defects (due to the C-H bonds) limit the electronic properties to lower mobilities compared to the QFS 1LG. Thus, the combination of X-ray photoelectron spectroscopy and surface enhanced Raman scattering provided a complete understanding of the underlying structure of quasi-free standing graphene and giving information of vibrational modes not observable using conventional Raman spectroscopy. Furthermore, the employment of Kelvin probe force microscopy provided detailed high-resolution maps of the surface potential distribution in vacuum and increase in the work function of graphene following the intercalation process, thus differentiating the graphene layers thicknesses. These observations conclude that the transformation of the IFL to QFS 1LG using hydrogen intercalation is the most promising route for the development of quasi-free standing graphene on SiC.

Author Contributions

O.K. designed the research and performed the SEM on Au particles, W.S. grew and intercalated the graphene, C.M., S.S., A.S. performed the measurements. All authors analyzed the results and participated in writing and reviewing the manuscript text.

Funding Sources

EC grants Graphene Flagship (No. CNECT-ICT-604391), EMRP under project GraphOhm (No. 117359) and NMS under the IRD Graphene Project (No. 119948). The work was carried out as part of an Engineering Doctorate program in Micro- and NanoMaterials and Technologies, financially supported by the EPSRC under the grant EP/G037388, the University of Surrey and the National Physical Laboratory.

Acknowledgments

The authors are especially grateful to Thomas Seyller and Florian Speck for the useful discussions and assistance with the XPS analysis. The authors are also grateful to Héctor Corte-León for developing the MATLAB scripts used in Raman map analysis.